\documentclass[journal]{IEEEtran}

\usepackage{comment}
\usepackage{cite}

\usepackage[cmex10]{amsmath}
\usepackage{amssymb}

\usepackage{algorithmic}
\usepackage{array}

\usepackage{stfloats}
\usepackage{url}

\usepackage{graphics}
\usepackage{epsfig}
\usepackage{graphicx}
\usepackage[tight,footnotesize]{subfigure}
\usepackage{epstopdf}

\title{Dynamic Modeling of Price Responsive Demand in Real-time Electricity Market: Empirical Analysis}

\author{
	Jaeyong An, P. R. Kumar, and~Le~Xie\\

	\thanks{
		This work is supported in part by NSF Contract ECCS-1546682,
		NSF Science \& Technology Center Grant CCF-0939370, 
		and the Power Systems Engineering Research Center (PSERC).
		
		The authors are with the Department of Electrical and Computer Engineering,
		Texas A\&M University, College Station, TX 77843 USA
		(e-mail: \texttt{jyan@tamu.edu, prk@tamu.edu, le.xie@tamu.edu} ).		
	}% <-this % stops a space
}

\begin{document}

\maketitle

\begin{abstract}
In this paper, we study the price responsiveness of electricity consumption from empirical commercial and industrial load data obtained from Texas.
Employing a dynamical system perspective, we show that price responsive demand can be modeled as a hybrid of a Hammerstein model with delay following a price surge, and a linear ARX model under moderate price changes.
It is observed that electricity consumption therefore has unique characteristics including
(1) qualitatively distinct response between moderate and extremely high prices; and 
(2) a time delay associated with the response to high prices.
It is shown that these observed features may render traditional approaches to demand response and retail pricing based on classical economic theories ineffective.  
In particular, ultimate real-time retail pricing may be limitedly beneficial than as considered in classical economic theories.
%In particular, ultimate real-time retail pricing may adversely affect the efficiency of the market.
\end{abstract}


\IEEEpeerreviewmaketitle

\section{Introduction} \label{sec:Intro}

\begin{figure}[!t]
\centerline{
	\includegraphics[width = 3 in]{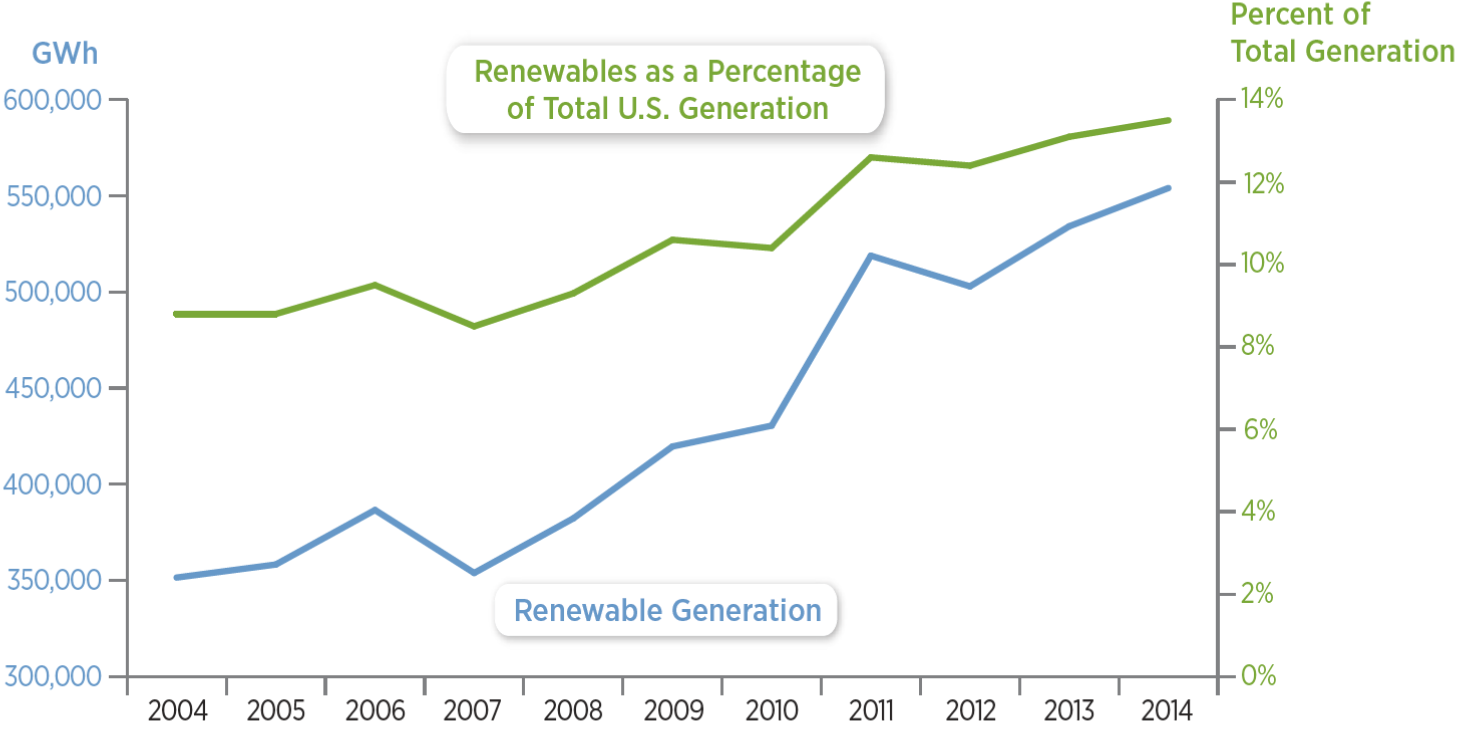}
}
\caption{The growth of renewable generation in the U.S. \cite{redb}}
\label{fig:RenewablePercentage}
\end{figure}

\IEEEPARstart{I}{n}, 
response to the emerging carbon emissions constrained world, 
the usage of renewable energy sources is increasing.
The overall increase in penetration of renewable energy resources in the U.S. is depicted in Fig. \ref{fig:RenewablePercentage}.
Such growth of renewable energy resource is not limited to the U.S.
Globally, installed global renewable electricity capacity has continued to increase and represented 28.5\% of total electricity capacity in 2014 \cite{redb} \cite{epa}. 

Renewable energy sources are generally characterized as
\emph{variable energy resources} (VER) due to their variability and uncertainty \cite{tfed}. 
While there has been efforts for better control of resources such as wind farms as doubly fed induction generators \cite{Hiskens},
their limited controllability and lack of predictability pose new challenges for the operation of the power system. 

Price-responsive demand, or \textit{demand response} (DR) is a key mechanism to achieve system balancing. One avenue is by modifying consumption patterns through economically exposing customers to time-varying pricing that reflect supply-demand balancing status.
A number of programs on DR have been implemented or proposed \cite{Albadi} \cite{Siano}.
While such price-responsive demand can potentially provide the key to system operability under high penetration of VERs, the presumed benefits of DR programs substantially depend on how responsive demand actually is to price: the price elasticity of demand \cite{Allcott11}.

Though price elasticity of demand is critical for the effectiveness of DR programs, 
previous empirical works based on data-driven \emph{static} analysis of demand suggests that even load labeled as price-responsive is fairly inelastic \cite{Xie}.
On the other hand, viewing DR as a \emph{dynamical system},
our previous work \cite{An} analyzing industrial and commercial loads of which is directly exposed to \emph{Electric Reliability Council of Texas} (ERCOT) real-time wholesale market makes two observations: 

\begin{enumerate}
  \item
The consumer's response to large prices (over the 95\%-quantile: \$144.42) can be modeled as a \emph{Hammerstein system},
i.e., a static nonlinearity followed by a linear transfer function \cite{Ding}.
After accounting for this nonlinear transformation,
which is typically concave since the response is sublinear,
the response exhibits a reduction after a delay of about 0.75-2.5 hours,
before subsequently reverting back to normal levels.

\item
The response to moderate prices (up to \$144.42) can be modeled as a linear stochastic system, specifically as an \emph{autoregressive exogenous} (ARX) system,
i.e., an autoregressive (AR) system with exogenous input and white noise.
\end{enumerate}

This paper extends our previous work \cite{An}, by analyzing the economic effect of characteristics of consumer behavior that prevent real-time retail electricity pricing from optimal signaling and respon.

The rest of this paper is organized as follows.
In Section \ref{sec:LitReview}, previous works analyzing the models and benefits of price responsive demand, mostly conducted in the economics literature, are reviewed.
We introduce our observations on consumer behavior from empirical load data from ERCOT in Section \ref{sec:ConsmObs}.
In Section \ref{sec:EffRTP}, on the basis of our empirical observation we discuss the implication of our observations, presenting an alternative analysis of the potential benefits of DR in comparison to previous literature.
Concluding remarks followed in Section \ref{sec:Conc}.

\section{Literature Review} \label{sec:LitReview}

While the idea of DR is currently attracting wide interest as a solution for system operability under high penetration of VERs, the necessity of DR has been advocated for decades by economists from a market efficiency perspective.
The volatility of load that has been challenge for system operators to cope with also entails abrupt and drastic changes in electricity price in the wholesale market.
Though extreme price fluctuation is widely observed in today's restructured electricity wholesale competitive markets, retail customers in most regions do not face frequent price change.
While wholesale electricity prices vary from hour to hour, retail prices do not change for months in most electricity markets.
Such discordance between rapid fluctuation in wholesale prices and near flat retail prices not only incurs economic inefficiency in terms of social welfare, but also creates price-inelastic wholesale demand that severely exacerbates the volatility of wholesale electricity prices. 
%In practice, wholesale electricity prices in real-time sometimes vary by an order of magnitude.
The combination of inelastic demand with the inherent real-time nature of the market makes electricity markets vulnerable to the exercise of market power \cite{Borenstein00}.

As a method to achieve price responsive demand,
there has been a consensus on the potential benefits of \emph{real-time retail pricing} (RTRP) among economists
  \cite{Joskow} \cite{Wolak} \cite{Borenstein02}  \cite{Borenstein03} \cite{Borenstein07} \cite{Hogan}.
%RTRP is one of the most important issues in the power industry, there has been an abundant literature supporting the economic benefits realizable from RTRP. 
The first potential benefit most discussed in the literature is the allocative efficiency improvement resulting from resolving the market inefficiency caused by (near) constant retail electricity prices 
%justified both via an econometric approach \cite{Aubin} and by theoretical analysis 
\cite{Allcott13} \cite{Borenstein03} \cite{Borenstein05b} \cite{Hogan} \cite{Aubin} \cite{Holland}.
The second benefit studied is the increased robustness of the market with RTRP forestalling the exercise of market power 
%\footnote{Market power is the ability of a firm to profitably raise the market price of a good or service over marginal cost mostly based on its own market share. The exercise of market power is an attempt by a firm to manipulate market price utilizing its market power. A critical condition for attaining the efficiency of a free competitive market is that every market participant is a price-taker. In that sense, a greater market power of a firm may bring a greater risk of market failure. If demands are inelastic, withholding a small portion of supply by a firm may drive the market price higher.}
\cite{Allcott12} \cite{Borenstein99} \cite{Borenstein00} \cite{Borenstein05a}.
The last benefit considered is that the mitigation of demand volatility induced by real-time price signals will also relieve the need for excessive reserve requirement which incurs a large portion of the societal costs
\cite{Allcott11} \cite{Allcott13} \cite{Joskow} .
However, all the potential economic benefits of RTRP substantially depend on how responsive demand is to price, i.e., the price elasticity of demand \cite{Allcott11} \cite{Allcott13}.

\begin{figure}[!t]
\centerline{
	\includegraphics[width = 2.5in]{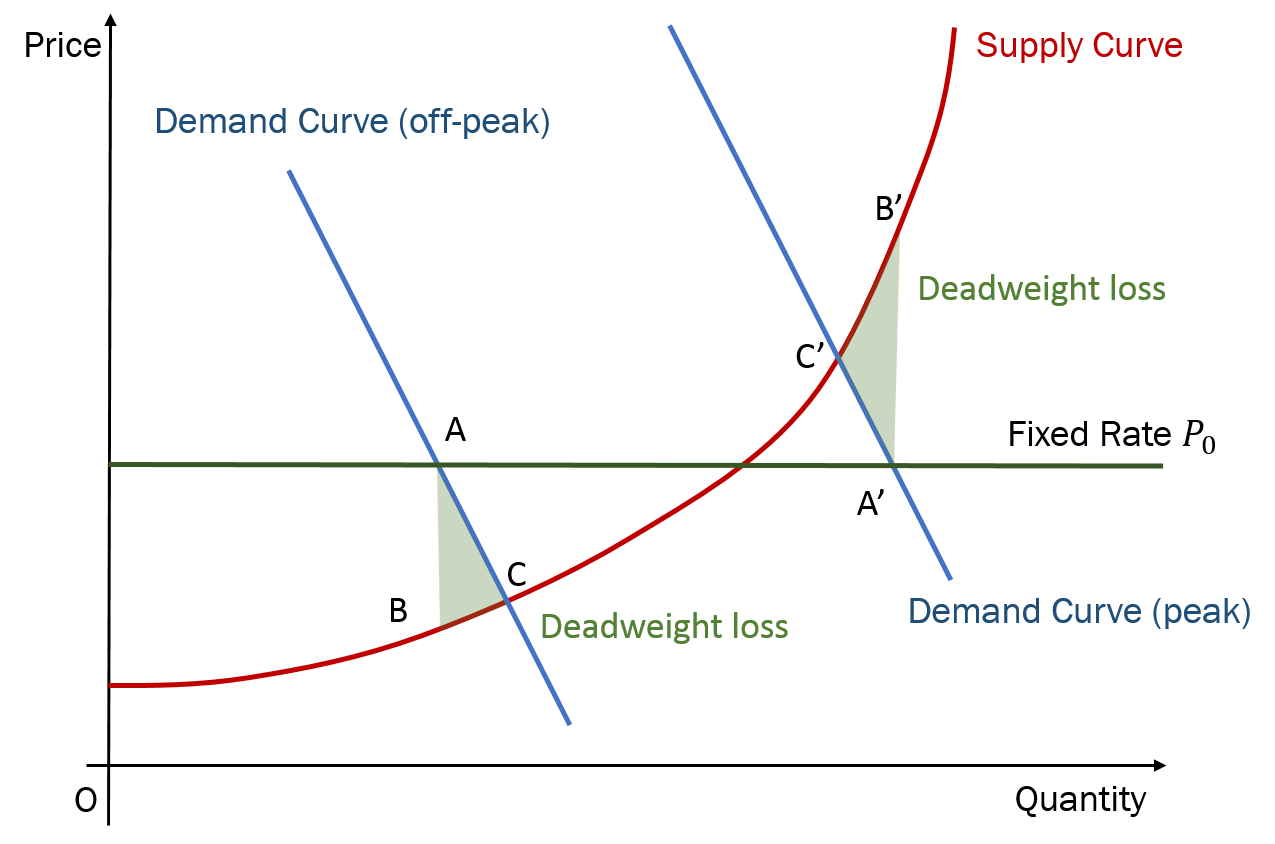}
}
\caption{The analysis of economic inefficiency under volatile demand resulting from a fixed retail electricity tariff.}
\label{fig:EconIneffFixRate}
\end{figure}

The efficiency improvement of RTRP is well analyzed in the literature \cite{Allcott13} \cite{Borenstein05b} \cite{Hogan} \cite{Holland}, as depicted in Figure~\ref{fig:EconIneffFixRate}.
Since the demand curve has a time variant property, it is not likely to happen that the fixed rate meets $C$ or $C'$, which are the optimal market clearing prices in terms of social welfare maximization. Thus, the shaded triangles $\Delta ABC$ and $\Delta A'B'C'$ are the deadweight loss, the economic inefficiency caused by the fixed rate $P_0$.
Due to the instantaneousness of electricity, it is reasonable to assume that the electricity at each time slot is a distinct commodity. 
%For instantaneous and accurate market convergence to the optimal market clearing prices ($C$ or $C'$ in Figure~\ref{fig:EconIneffFixRate}), 
Thus, RTRP advocates argue that the ultimate real-time retail price is the optimal pricing policy\cite{Borenstein05b} in terms of economic efficiency.

Although the analysis shown in Figure~\ref{fig:EconIneffFixRate} seems reasonable, it requires a crucial assumption to be justified: 
\emph{Demand converges to $C$ or $C'$ almost immediately, in at most one time slot as determined by the market rules.} 
%and (2) \emph{Utility from the consumption in each time slot is attained in that same time slot.} 
However, this assumption is controversial 
when the market is fast-paced.
%unless the market is slow-paced. 
%However, the electricity consumption for running a laundry machine (say) does not provide utility until the laundering is complete, which may take more than one period.
Additionally, a fundamental limitation in the demand-supply curve model is that it is difficult to obtain any insight concerning dynamic behavior from the demand curve, which makes it difficult to estimate and predict demand from this static model.
The goal of our work is to develop a model for demand response as a stochastic dynamical
system where both past prices and past consumption influence future consumption probabilistically.

%The deadweight loss inefficiency of any price fixed \textit{ex ante} is proportional to the square of the deviation from real-time price.

\section{Empirical Observations on Consumers' Behavior in ERCOT} \label{sec:ConsmObs}

In this section, we introduce our prior work \cite{An} that poses the problem of modeling price responsive demand at wholesale level. 
This work is based on an analysis of the data from an anonymous commercial/industrial (C/I) load\footnote{Anonymous even to us.} in Houston, purchasing its power directly from ERCOT real-time wholesale market, gleaned over nine months (Jan.1 - Sep. 30, 2008).
Based on the empirical data, a dynamical model of consumer behavior is presented.

\subsection{Preliminary Data Analysis} \label{par:Prlim}

\begin{figure}%[!t]
\centerline{
    \subfigure[The boxplot of hourly load.]{
        \includegraphics[width = 1.8in]{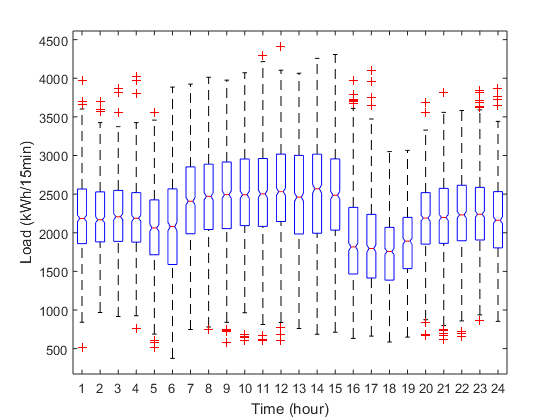}
        \label{fig:Load-boxplotHourly}
    }
    \subfigure[The boxplot of hourly prices.]{
        \includegraphics[width = 1.8in]{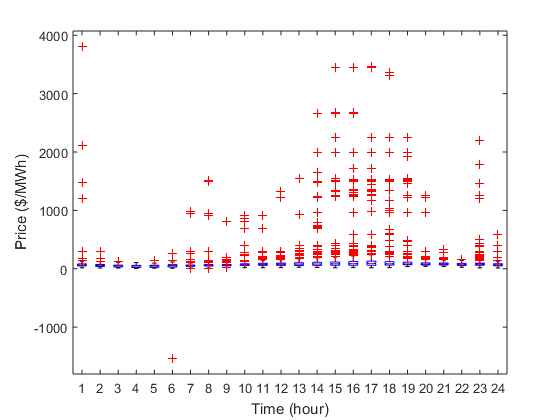}
        \label{fig:Price-boxplotHourly}
    }
}
\centerline{
    \subfigure[The median price by time of day (at 15-minute intervals).]{
        \includegraphics[width = 1.8in]{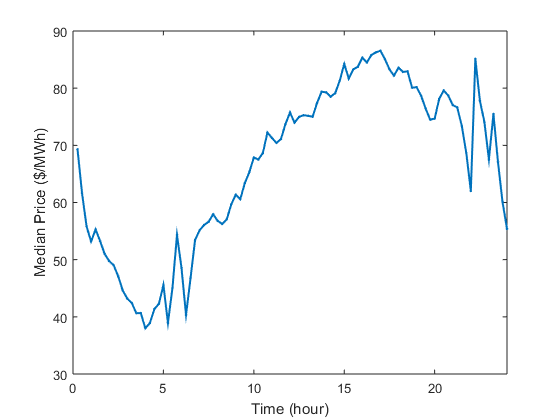}
        \label{fig:Price-medianHourly}
    }
    \subfigure[The price time series on June 8.]{
        \includegraphics[width = 1.8in]{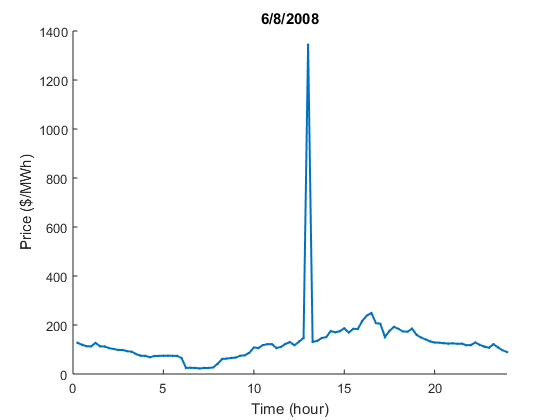}
        \label{fig:Price-Sample}
    }
}
\caption{Figs.  \ref{fig:Load-boxplotHourly} and \ref{fig:Price-boxplotHourly} show the hourly plots of a C/I load and prices from ERCOT, based on 15-minute measurements from Jan. 1, 2008 to Sep. 30, 2008. Fig. \ref{fig:Price-medianHourly} shows the median price over these nine months by time of day. Fig. \ref{fig:Price-Sample} shows a particular sample of the price series on June 8, 2008.}
\label{fig:PQ-HourlyPlot}
\end{figure}

The C/I load and prices from Houston measured at intervals of 15 minutes from Jan. 1, 2008 to Sep. 30, 2008 is presented with respect to time in Fig. \ref{fig:PQ-HourlyPlot}. 
The first notable point observed here is that the plot on price (Fig. \ref{fig:Price-boxplotHourly}) shows many outliers while the plot on load rarely has any.
This is called the ``spiky" nature of electricity prices, an irregular sudden extreme price change for a very short duration of 15-30 minutes (Fig. \ref{fig:Price-Sample}). This gives the price a highly non-normal heavy-tail distribution. 
The fundamental reason for the spiky nature of prices is explained in Section \ref{sec:LitReview}.
The other property we see in Fig. \ref{fig:PQ-HourlyPlot} is that the time-series of load shows a depressed demand in ``peak hours'' (afternoons), over time intervals
that overlap with the time intervals exhibiting frequent large outliers in the price time series.
Here, we surmise that the depressed demand is a manifestation of demand response, and 
that this demand response is highly connected to the outliers of price,
because the depression is not likely to be explained by the plot of the median prices (Fig. \ref{fig:Price-medianHourly}).

\begin{figure}%[!t]
\centerline{
%    \subfigure[The histogram of demand ($Q$).]{
%        \includegraphics[width = 1.8in]{figs/Load-hist}
%        \label{fig:Load-hist}
%    }
    \subfigure[The cumulative probability distribution of price ($P$) versus that of the normal distribution.]{
        \includegraphics[width = 1.8in]{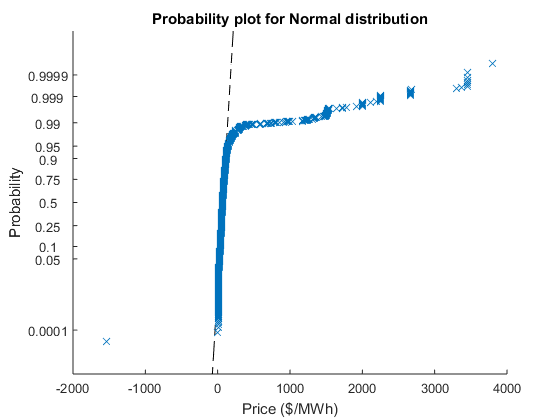}
        \label{fig:Price-probplot}
    }
    \subfigure[The empirical cumulative probability distribution of the demand ($Q$) and the comparison with the normal distribution.]{
        \includegraphics[width = 1.8in]{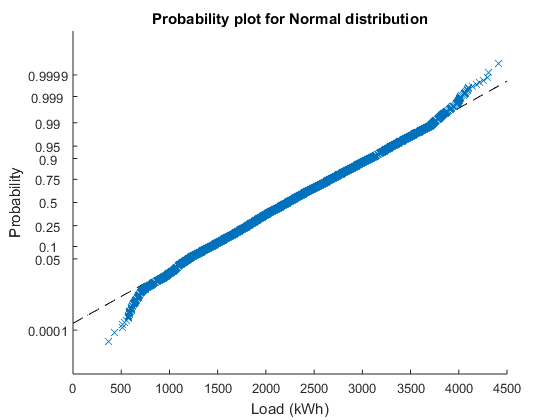}
        \label{fig:Load-probplot}
    }
}
\centerline{
    \subfigure[The autocorrelation function (ACF) of $Q$. (One discrete unit of time = 15 mins)]{
        \includegraphics[width = 1.8in]{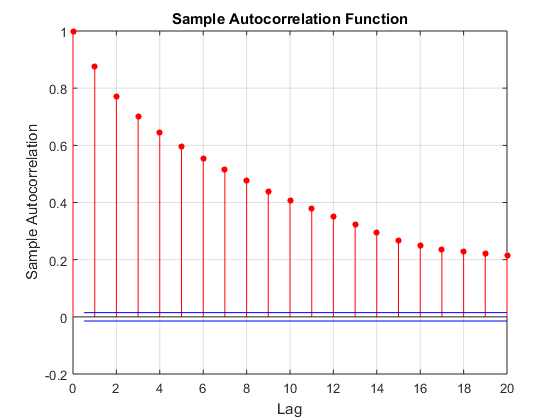}
        \label{fig:Load-autocorr}
    }
    \subfigure[The partial autocorrelation function (PACF) of $Q$ (The autocorrelation of each lag $k$ after the dependence on lags $1, 2, \ldots, k-1$ is removed).]{
        \includegraphics[width = 1.8in]{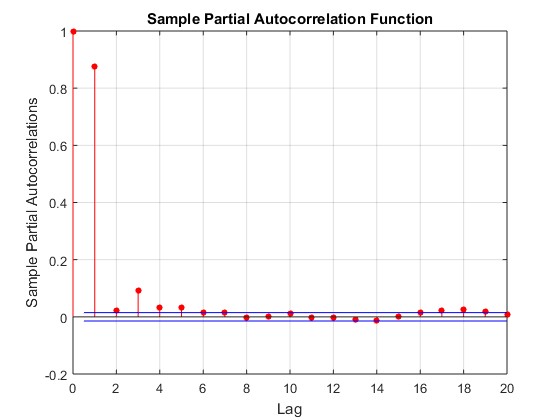}
        \label{fig:Load-parcorr}
    }
}
\caption{The statistics of Price ($P$) from ERCOT and the C/I load ($Q$) on workdays (i.e., weekends removed) based on 15-minute measurements from Jan. 1, 2008 to Sep. 30, 2008.}
\label{fig:PrelimStat}
\end{figure}

\begin{table}%[!t]
\renewcommand{\arraystretch}{1.3}
\caption{Statistics of price (P) and load ($Q$)}
\label{tab:Preliminary}
\centering
\begin{tabular}{ccccc}
\hline
& \bf{Kurtosis} & \bf{Skewness} & \bf{Mean} & \bf{Std. Deviation}\\
\hline
P & 149.0002 & 10.9133 & 67.9700 & 173.2434\\
\hline
Q & 2.7712 & 0.1069 & 2246 & 631.0869\\
\hline
\end{tabular}
\end{table}

In Fig. \ref{fig:PrelimStat}, the statistics of the prices (P) and C/I load ($Q$) on workdays are shown.
%From Fig. \ref{fig:Load-hist}, we can see that the empirical distribution of the load is fairly close to the normal distribution.
%This is clearer in the empirical probability plot of load versus the normal distribution (Fig. \ref{fig:Load-probplot}) shown by the diagonal dashed line.
In Fig. \ref{fig:Load-probplot}, the empirical probability plot of load versus the normal distribution is indicated by the diagonal dashed line, and
we can check that this empirical distribution of the load can be assumed to be a normal distribution.
For further validation, we can also check an estimate of the kurtosis, $\mu_4 / \sigma^4$, where $\mu_n$ is the $n$th moment about the mean and $\sigma$ is the standard deviation. 
It is 2.77, which is close to the value 3.0 for the normal distribution. 
Also, its skewness, $\mu_3 / \sigma^3$, is 0.11, which is close to the value 0 for the normal distribution. (Table \ref{tab:Preliminary}). Therefore, we can conclude that the distribution of the load is near normal.
In Fig. \ref{fig:Load-autocorr}, the plot of autocorrelation (ACF) of the load shows a high correlation between the current and the past load,  while the partial autocorrelation (PACF) of the load shown in Fig. \ref{fig:Load-parcorr} decays rapidly in no more than five quarter hours (75 minutes). Taking these facts into account, a simple autoregressive (AR) model of order 3 or 5 is concluded to sufficiently well describe the load process.

On the other hand, the first feature we can see in Fig. \ref{fig:Price-probplot} is that the distribution of prices is highly non-normal.
The cumulative distribution matches the diagonal dashed line,
suggesting closeness to the normal distribution at low to moderate prices.
However, the top 5\% of the prices deviate severely from the line, reflecting the spiky nature of electricity prices.
Such a long-tail property yields huge kurtosis (149.0002) and skewness (10.9133) as shown in Table \ref{tab:Preliminary}.

From the above, it is obvious that it is not feasible to find a linear relationship between load and price over all values of $P$ and $Q$.
Hence, we conclude that it is not possible to obtain one single all encompassing universal linear dynamic
system model between price and demand.
As an alternative, we continue the analysis by assuming that
there are \emph{two transfer functions} (TFs), one for \emph{moderate prices} which is a linear model,
and one for \emph{high prices} which is nonlinear.
The deviation from normality of the top 5\% in Fig. \ref{fig:Price-probplot} provides a reasonably good demarcation between moderate prices and high prices.

\subsection{Estimation of Dynamic model on Load and Price} \label{par:DR}

From the preliminary data analysis in Section \ref{par:Prlim}, 
we infer that there exist two qualitatively distinct regimes,
a moderate price regime, and a high price regime. 
In the former, we consider a linear transfer function between price and load, with additional noise to account for uncertainty, i.e., an ARX model driven by white noise.
In the high price regime, we consider a concave transformation of peak prices
to account for the non-normality of the process.
In this section, we further address the problem of the dynamic model identification of DR.

\subsubsection{Methodology} \label{par:DR-method}

We briefly discuss the estimation and validation methodology for the estimation of the dynamic model of DR.
As a dynamic model of DR, we consider an ARX model driven by white noise, 
one of the simplest but most utilizable models for forecasting and control.
For estimation, we use the least squares (LS) method for estimating the unknown parameters of a linear regression model \cite{Hayashi}, \cite{Kumar}.
For the verification of the existence of DR and the significance of the results of estimated parameters, we consider the \emph{analysis of variance} (ANOVA) method \cite{Fisher}.
For examining the minimum net contribution of price information to load estimation, we conduct a two-step estimation procedure.
To achieve parsimony of the model, we cross-validate the model by a random division of each complete data set under the two separate conditions (i.e., moderate prices and high prices) into two sets of the same size, namely, a training set for estimation and a test set for evaluation.

\subsubsection{Autoregressive Exogenous (ARX) model}

Denote by $\{P(t)\}^N_{t=1}$ and $\{Q(t)\}^N_{t=1}$ the time series of prices and loads, each consisting of $N$ observations.
Denoting by $z^{-1}$ the backwardshift operator $z^{-1}X(t) := X(t-1)$,
the ARX model can be represented as follows:
\begin{equation}    \label{eq:ARX-C}
\alpha(z^{-1})Q(t) = \beta(z^{-1})P(t) + \epsilon_t,
\end{equation}
where vectors
$\alpha := [1 \ -\alpha_1 \ -\alpha_2 \ ... \ -\alpha_m]'$ and
$\beta := [\beta_1 \ \beta_2 \ ... \ \beta_n]'$ are unknown parameters to be estimated,
$\alpha(z^{-1}) :=  \alpha' \cdot [z^{-i}]_{i=1}^m$ and
$\beta(z^{-1}) := \beta' \cdot [z^{-i}]_{i=1}^n$ are the characteristic and numerator polynomials of the TF respectively,
 and
$\epsilon_t$ is an error which is an independent and identically distributed (i.i.d.) noise process with expectation $E\epsilon_t = 0$ and variance $\text{VAR} \epsilon_t = \sigma^2$.

\paragraph{Two-step Estimation}
%Our objectives are not only to forecast load via price but also to show the existence of DR and understand it.
Our primary objective is to determine the existence of DR, and understand it, if it exists, from a dynamic system perspective.
We employ the following two-step estimation procedure to examine the net contribution of price to load.

\begin{enumerate}
  \item First estimate the regression parameters $\mathbf{\hat{\alpha}}$, and obtain $Q_{res}(t) := (1 - \sum_{i=1}^m{\hat{\alpha_i} z^{-i}) Q(t)}$.
  \item Estimate $\mathbf{\hat{\beta}}$ using the equation $Q_{res}(t) = (\sum_{i=1}^n{\beta_i z^{-i}}) P(t) + \epsilon_t$.
\end{enumerate}

Then, the overall estimated ARX model is the following:
\begin{equation}
Q(t) = (\sum_{i=1}^m{\hat{\alpha_i} z^{-i}}) Q(t) + (\sum_{i=1}^n{\hat{\beta_i} z^{-i}}) P(t) + \epsilon_t,
\end{equation}
where $\hat{\alpha_i}$ and $\hat{\beta_i}$ are the LS estimators of $\alpha_i$ and $\beta_i$.

\subsection{Demand Response to Moderate Price} \label{par:DR-low}

In this section, an ARX model for DR in the moderate price regime, the prices below the 95\%-quantile, is presented.
The overall estimation result of fitting an ARX model are shown in Tables \ref{tab:DR-Moderate3}.
The estimated TF of the model is:
\begin{equation}    \label{eq:TF-Low}
TF_{\text{Low}} = \frac{-0.8555 z^{-1} + 0.5273 z^{-2}}{1 - 0.8127 z^{-1} - 0.0461 z^{-3} - 0.0366 z^{-5}}.
\end{equation}

Tables \ref{tab:DR-Moderate1} and \ref{tab:DR-Moderate2} present the
results of the analysis for each of the two steps of estimation.
The \emph{Estimate} column shows the estimated coefficient value,
\emph{SE} refers to the standard error of the estimate,
\emph{tStat} indicates the t-statistic for a hypothesis test that the coefficient is zero, and
\emph{pValue} is the p-value for the t-statistic.
This model explains 77.6\% of the variance that $Q(t)$ initially possesses.

\begin{table}%[!t]
\renewcommand{\arraystretch}{1.3}
\caption{Estimated AR Model of $Q(t)$}
\label{tab:DR-Moderate1}
\centering
\begin{tabular}{ccccc}
\multicolumn{5}{c}{$Q(t) = \alpha_1 Q(t-1) + \alpha_3 Q(t-3) + \alpha_5 Q(t-5) + \alpha_0 + Q_{res}(t)$}\\
\hline
\bf{Coeff.} & \bf{Estimate} & \bf{SE}	& \bf{tStat} & \bf{pValue}\\
\hline
$\alpha_0$ & 238.07 & 13.989 & 17.018 & $8.883 \times 10^{-64}$\\
$\alpha_1$ & 0.81268 & 0.0085477 & 95.075 & 0\\
$\alpha_3$ & 0.046086 & 0.010267 & 4.4886 & $7.2744 \times 10^{-6}$\\
$\alpha_5$ & 0.036614 & 0.0085466 & 4.284 & $1.8579 \times 10^{-5}$\\
\hline

\multicolumn{4}{c}{$\sqrt{\text{MSE}}$ : 301} &
\multicolumn{1}{c}{$R^2$: 0.775}    \\
\multicolumn{4}{c}{F-statistic vs. constant model: $8.81 \times 10^3$} &
\multicolumn{1}{c}{p-value = 0} \\
\hline
\end{tabular}
\end{table}

Though price has sufficient statistical significance due to its low p-value (0.0147), 
what we see here is that its innovative contribution to the load forecast is relatively small (less than 0.1\%), and most of the change in $Q(t)$ can be explained by the past of the load itself (AR(5) model).
This suggests that a moderate price has very little impact in terms of eliciting demand response. 
This is also consistent with our observation in the preliminary analysis in Section \ref{par:Prlim}.

\begin{table}%[!t]
\renewcommand{\arraystretch}{1.3}
\caption{Estimated Linear Model of $Q_{res}(t)$}
\label{tab:DR-Moderate2}
\centering
\begin{tabular}{ccccc}
\multicolumn{5}{c}{$Q_{res}(t) = \beta_1 P(t-1) + \beta_2 P(t-2) + \beta_0 + \epsilon_t$}\\
\hline
\bf{Coeff.} & \bf{Estimate} & \bf{SE}	& \bf{tStat} & \bf{pValue}\\
\hline
$\beta_0$ & 22.506 & 10.054 & 2.2385 & 0.025218\\
$\beta_1$ & -0.8555 & 0.42677 & -2.0046 & 0.045043\\
$\beta_2$ & 0.5273 & 0.43006 & 1.2261 & 0.2202\\
\hline
\multicolumn{3}{c}{$\sqrt{\text{MSE}}$ : 301} &
\multicolumn{2}{c}{$R^2$: 0.00084}\\
\multicolumn{3}{c}{F-statistic vs. constant model: 4.22} &
\multicolumn{2}{c}{p-value = 0.0147}\\
\hline
\end{tabular}
\end{table}

\begin{table}%[!t]
\renewcommand{\arraystretch}{1.3}
\caption{The ARX Model on $Q(t)$}
\label{tab:DR-Moderate3}
\centering
\begin{tabular}{cccc}
%\multicolumn{4}{l}{$(1 - \alpha_1z^{-1} - \alpha_3z^{-3} - \alpha_5z^{-5})Q(t) = (\beta_1z^{-1} + \beta_2z^{-2}) P(t) + \epsilon_t + \epsilon_0$}\\
\multicolumn{4}{l}{$(1 - \alpha_1z^{-1} - \alpha_3z^{-3} - \alpha_5z^{-5})Q(t)$ \hspace{1.5 cm}} \\
\multicolumn{4}{r}{\hspace{1.5 cm} $= (\beta_1z^{-1} + \beta_2z^{-2}) P(t) + \epsilon_t + \epsilon_0$}\\
\hline
\bf{Coeff.} & \bf{Estimate} & \bf{Coeff.} & \bf{Estimate}\\
\hline
$\alpha_1$ & 0.81268 & $\beta_1$ & -0.8555\\
$\alpha_3$ & 0.046086 & $\beta_2$ & 0.5273\\
$\alpha_5$ & 0.036614 & $\epsilon_0$ & 260.126\\
\hline
\multicolumn{2}{c}{$\sqrt{\text{MSE}}$ : 301} &
\multicolumn{2}{c}{$R^2$: 0.776}\\
%$\sqrt{\text{MSE}}$ : & 301 & $R^2$: & 0.776\\
\hline
\end{tabular}
\end{table}

\subsection{Demand Response to High Price} \label{par:DR-high}

\begin{figure}[!t]
\centerline{
    \subfigure[The sample series of load change in response to the price spike at 3:30pm Apr. 3, 2008.]{
        \includegraphics[width = 1.8in]{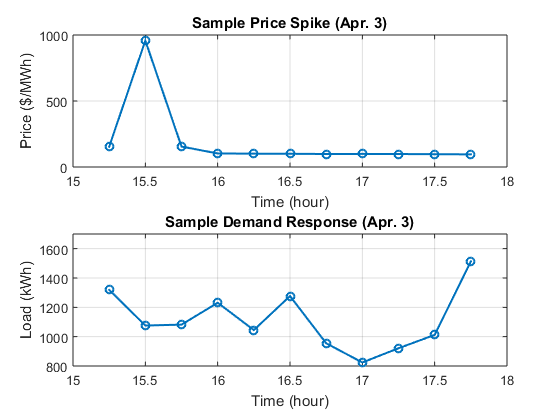}
        \label{fig:DR-sample}
    }
    \subfigure[The box plot of Q after a price surge (over 95\%-quantile) at lag=0.]{
        \includegraphics[width = 1.8in]{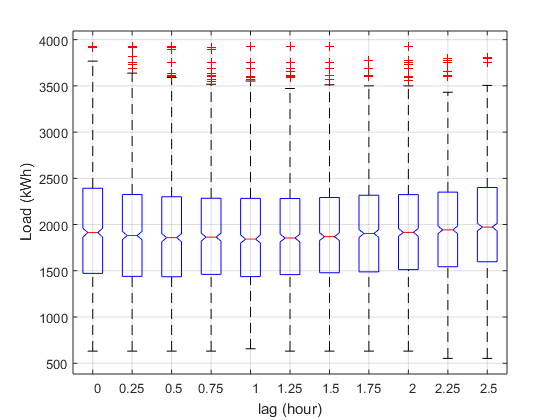}
        \label{fig:DR-anova1}
    }
}
\centerline{
    \subfigure[The average change in Q after different levels of price surges.]{
        \includegraphics[width = 1.8in]{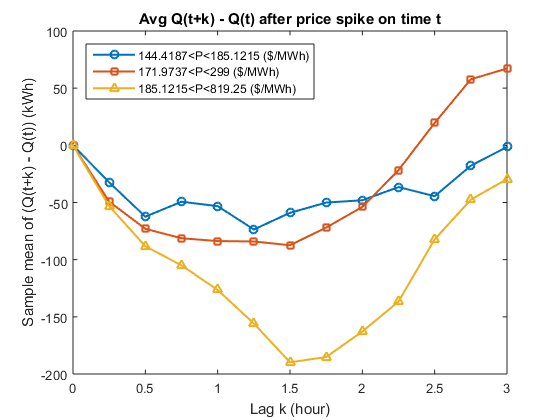}
        \label{fig:DR-Comparison}
    }
    \subfigure[The correlation between $\Delta P$ and $Q(t+k)$ after a price surge.]{
        \includegraphics[width = 1.8in]{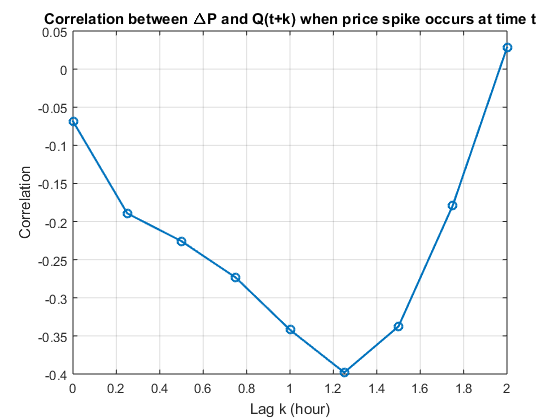}
        \label{fig:DR-Correlation}
    }
}
\caption{The temporal pattern of the change of Q in response to price surge}
\label{fig:DR-Example}
\end{figure}

In this section, an ARX model for the high price regime, 
where the prices are over the 95\%-quantile (144.4187 \$/MWh), is presented.
A sample load evolution time-series after a high price spike is shown in Fig. \ref{fig:DR-sample}.
What we can notice here is a huge drop of the load after a one and half hour lag. 
Fig. \ref{fig:DR-anova1} indicates that such a load drop phenomenon is not an isolated event; 
we commonly see such a load drop and recovery pattern over two and half hours after price surges. The ANOVA result of Fig. \ref{fig:DR-anova1} in Table \ref{tab:DR-anova1}, showing its extremely low p-value $(3.86 \times 10^{-4})$, supports our observation that there exists a significant load drop 0.5-1.5 hours after a price surge.
This significance level is sufficiently low to reject the null hypothesis of a price unresponsive model for the load.

In addition, we also see from Fig. \ref{fig:DR-Comparison} and \ref{fig:DR-Correlation} that the height of the price surges is correlated to the depth of load drop.
Fig. \ref{fig:DR-Comparison} depicts the average change
$\overline{Q(k)}$, at a certain level of price surge $P$ at time $t$, where
$\overline{Q(k)} := \frac{1}{|\bf{P}|} \sum_{P(t) \in \bf{P}}{[Q(t+k) - Q(t)]}$ for
all $\bf{P}$ in a subset of sample prices $\mathbf{P} = \{P(t): P_{\text{min}} \leq P(t) \leq P_{\text{max}} \}$ for given $P_{\text{min}}$ and $P_{\text{max}}$.
We check that higher $P_{\text{min}}$ and $P_{\text{max}}$ result in a greater load drop.
The negative correlation between the height of the price surge ($\Delta P = P(t)-P(t-1)$) and the load Q is plotted in Fig. \ref{fig:DR-Correlation}, 
which is negatively significant after $k=5$ quarter-hour periods (i.e., one hour and 15 minutes) following a price surge.

\begin{table}%[!t]
\renewcommand{\arraystretch}{1.3}
\caption{ANOVA Results for Fig. \ref{fig:DR-anova1}}
\label{tab:DR-anova1}
\centering
\begin{tabular}{ccccccc}
\hline
\bf{Source} & \bf{SS} & \bf{DF}	& \bf{MS} & \bf{F} & \bf{p-value}\\
\hline
Groups & $1.21 \times 10^7$ & 10 & $1.21 \times 10^6$ & 3.21 & $3.86 \times 10^{-4}$\\
Error & $3.89 \times 10^9$ & 10351 & $3.76 \times 10^5$\\
Total & $3.90 \times 10^9$ & 10361 & \\
\hline
\end{tabular}
\footnotemark{\\SS: Sum of squares; DF: Degree of freedom of error; \\ MS: Mean square; F: F-statistic.}
\end{table}

On the basis of the above observations, we establish a simple dynamic model between the magnitude of the price surge and the load, in the case of high price surges.
Taking into account the long-tailed characteristic of prices, we consider a linear model for a  concave transformation $\log P(t)$, instead of $P(t)$. 
In this paper, we present a TF for a specific time period, from 2:00pm to 2:30pm, due to the innate time-dependency on DR.
The estimation results for the ARX model of DR at high price are presented in Tables \ref{tab:DR-Peak1}, \ref{tab:DR-Peak2}, and \ref{tab:DR-Peak3}.
The estimated TF of the ARX model is:
\begin{equation}    \label{eq:TF-Peak}
TF^{2:15pm}_{\text{Peak}} = \frac{-220.1 z^{-4}}{1 - 0.4015 z^{-1} + 0.2383 z^{-2} - 0.2512 z^{-4}},
\end{equation}
which accounts for 51.2\% of the variance of $Q(t)$.
The notable feature we find here is that the accuracy of the AR model for $Q(t)$ is severely degraded ($R^2 = 33.2\%$) in Table \ref{tab:DR-Peak1}, compared to the AR model for the moderate price regime (Table \ref{tab:DR-Moderate1}).
However, we observe that a relatively high portion (27\%) of the variance of $Q_{res}(t)$ is explained by the estimated model of $Q_{res}(t)$ shown in Table \ref{tab:DR-Peak2}, from which we conclude that the innovation from the price information is significant to improve $R^2$ of the ARX model up to 51.2\%, as presented in Table \ref{tab:DR-Peak3}.

\begin{table}%[!t]
\renewcommand{\arraystretch}{1.3}
\caption{Estimated AR Model for $Q(t)$}
\label{tab:DR-Peak1}
\centering
\begin{tabular}{ccccc}
\multicolumn{5}{c}{$Q(t) = \alpha_1 Q(t-1) + \alpha_2 Q(t-2) + \alpha_4 Q(t-4) + \alpha_0 + Q_{res}(t)$}\\
\hline
\bf{Coeff.} & \bf{Estimate} & \bf{SE}	& \bf{tStat} & \bf{pValue}\\
\hline
$\alpha_0$ & 748.26 & 233.72 & 3.2015 & 0.0025097\\
$\alpha_1$ & 0.40153 & 0.11763 & 3.4133 & 0.0013678\\
$\alpha_2$ & -0.23826 & 0.1461 & -1.6308 & 0.10992\\
$\alpha_4$ & 0.25124 & 0.11516 & 2.1816 & 0.0344\\
\hline
\multicolumn{3}{c}{$\sqrt{\text{MSE}}$ : 336} &
\multicolumn{2}{c}{$R^2$: 0.332}\\
\multicolumn{3}{c}{F-statistic vs. constant model: 7.44} &
\multicolumn{2}{c}{p-value = 0.000377}\\
\hline
\end{tabular}
\end{table}

\begin{table}%[!t]
\renewcommand{\arraystretch}{1.3}
\caption{Estimated Linear Model for $Q_{res}(t)$}
\label{tab:DR-Peak2}
\centering
\begin{tabular}{ccccc}
\multicolumn{5}{c}{$Q_{res}(t) = \beta_4 logP(t-4) + \beta_0 + \epsilon_t$}\\
\hline
\bf{Coeff.} & \bf{Estimate} & \bf{SE}	& \bf{tStat} & \bf{pValue}\\
\hline
$\beta_0$ & 1213.4 & 293.68 & 4.1316 & 0.00014688\\
$\beta_4$ & -220.1 & 52.774 & -4.1707 & 0.00012965\\
\hline
\multicolumn{3}{c}{$\sqrt{\text{MSE}}$ : 281} &
\multicolumn{2}{c}{$R^2$: 0.27}\\
\multicolumn{3}{c}{F-statistic vs. constant model: 17.4} &
\multicolumn{2}{c}{p-value = 0.00013}\\
\hline
\end{tabular}
\end{table}

\begin{table}%[!t]
\renewcommand{\arraystretch}{1.3}
\caption{The ARX Model for $Q(t)$}
\label{tab:DR-Peak3}
\centering
\begin{tabular}{cccc}
\multicolumn{4}{c}{$(1 - \alpha_1z^{-1} - \alpha_2z^{-2} - \alpha_4z^{-4})Q(t) = \beta_4z^{-4} logP(t) + \epsilon_t + \epsilon_0$}\\
\hline
%\multicolumn{4}{l}{$(1 - \alpha_1z^{-1} - \alpha_2z^{-2} - \alpha_4z^{-4})Q(t)$ \hspace{1.5 cm}} \\
%\multicolumn{4}{r}{\hspace{1.5 cm} $= \beta_4z^{-4} logP(t) + \epsilon_t + \epsilon_0$}\\
%\hline
\bf{Coeff.} & \bf{Estimate} & \bf{Coeff.} & \bf{Estimate}\\
\hline
$\alpha_1$ & 0.40153 & $\beta_4$ & -220.1\\
$\alpha_2$ & -0.23826 & $\epsilon_0$ & 1961.66\\
$\alpha_4$ & 0.25124\\
\hline
\multicolumn{2}{c}{$\sqrt{\text{MSE}}$ : 281} &
\multicolumn{2}{c}{$R^2$: 0.5124}\\
\hline
\end{tabular}
\end{table}

\begin{figure}%[!t]
\centerline{
    \subfigure[The probability plot of $\epsilon$ for normal distribution (Kurtosis = 3.1809).]{
        \includegraphics[width = 1.8in]{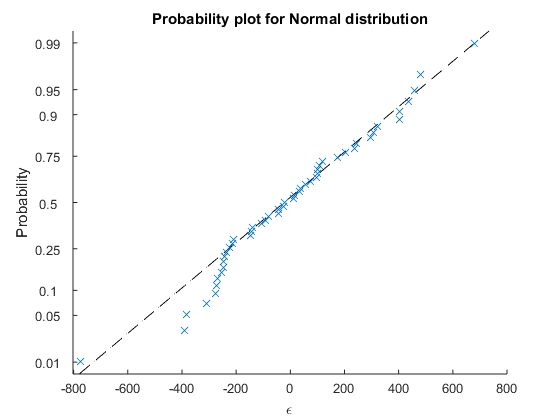}
        \label{fig:DR-Error-probplot}
    }
    \subfigure[The plot of $\hat{Q}$ over Q ($r_{\hat{Q}Q} = 0.7160$).]{
        \includegraphics[width = 1.8in]{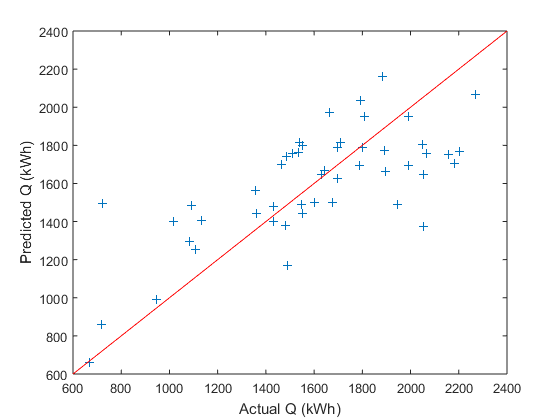}
        \label{fig:DR-QQhat}
    }
}
\caption{The plots of prediction error $\epsilon$.}

\label{fig:DR-Prediction}
\end{figure}

In Fig. \ref{fig:DR-Prediction}, The validity of our model is shown by sample load forecast.
Figs. \ref{fig:DR-Error-probplot} and \ref{fig:DR-QQhat} delineate the errors in the load forecast at 3:15pm after a price surge at 2:15pm.
We see that the forecasted $\widehat{Q}(t)$ and the actual $Q(t)$ at $t=\text{3:15pm}$ are reasonably well correlated (correlation ($r_{\hat{Q}Q} = 0.7160$) in Fig. \ref{fig:DR-QQhat}, and that the errors exhibit normality (Kurtosis = 3.1809) in Fig. \ref{fig:DR-Error-probplot}.

\subsection{Summary} \label{par:Summ}

Our empirical study suggests that 
(1) \emph{the demand only responds to high price surges at peak hour} and 
(2) \emph{there exists a demand response delay consequent on a high price surge.} 
The second finding shows that there exists a certain ``inertia'' in consumption, resulting in a certain time delay to reduce power consumption after a peak price observation.

\section{The Evaluation of the Benefit of Demand Exposure to Real-time Pricing} \label{sec:EffRTP}

In this section,
we shall further analyze the results from the data.
We will first examine the rationality of consumer behavior.
Then we will show that the observed delay in demand response changes classical arguments about the role of prices and the equilibrium process, as well as classical efficiency results of markets.

\subsection{Is the Observed Consumer's Behavior Rational?}
\begin{figure}%[!t]
\centerline{
    \subfigure[3:00 AM - 9:00 AM.]{
        \includegraphics[width = 1.8in]{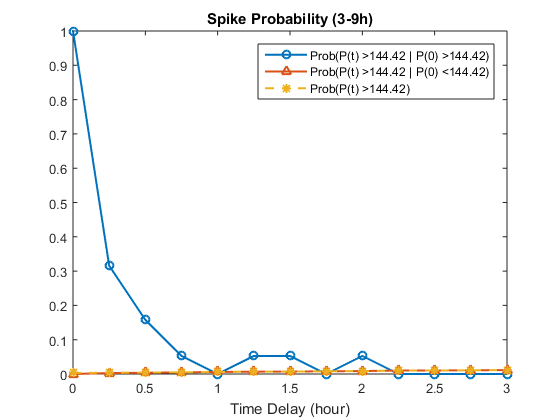}
        \label{fig:SpikeProb2}
    }
    \subfigure[9:00 AM - 3:00 PM.]{
        \includegraphics[width = 1.8in]{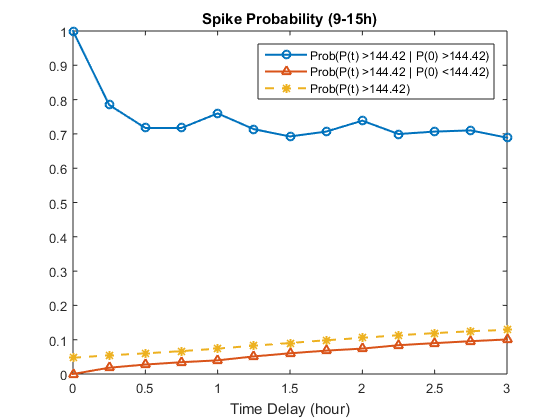}
        \label{fig:SpikeProb3}
    }
}

\caption{The comparison of estimated conditional probability of price spike occurrence after a time delay (horizontal axis) given a price within different price ranges.}
\label{fig:SpikeProb}
\end{figure}

One of our observations in Section \ref{sec:ConsmObs} is that
price responsive demand exhibits delayed response to price shock at peak hours.
It may seem to be irrational to decrease one's demand after a price shock has already occurred.
However, if we consider that consumer behavior is based on \emph{prediction} of price, rather than the current price itself, then we can explain the delayed response based on the inertia of demand.
In this sense, decreasing one's demand after a price spike, specifically, during or after the price plummets after price surge, can be well explained as a rational behavior if there is a high chance of a price increase after price spike. 
The chance of such a price increase relapse after a price spike is presented in Figure \ref{fig:SpikeProb}.

Figure \ref{fig:SpikeProb} shows a comparison of the estimated conditional probability of a high price in different situations, based on the obtained data from Houston. 
We can easily discern from Figure \ref{fig:SpikeProb} that 
the conditional probability of a second price spike following the occurrence of a price spike quickly reduces in off-peak hours. 
However, we also observe that the conditional probability of a price spike after the occurrence of a price spike during peak-hours remains at a significantly higher level than the probability of price spike without any conditioning.
This rationalizes our observation that \textit{if we assume a consumer has limited ability for immediate load reduction}, then a rational consumer adjusts its load in response to price spike in spite of its inertia, because the relative chance of repeated price surge after a price spike is significantly high, and the demand is still not able to respond quickly to that following price surge.
This also explains why demand is responsive to price only when it is during peak hours.
The rigorous analysis of the consumer rationality behind such behavior will be presented in subsequent work.

\subsection{Potential Modification of Classical Arguments About Price Responsive Demand}

Based on the assumed rationality of a consumer, 
the premise that demand is an optimal choice in consumption space given a price bundle under individual's budget constraint, is the major basis for classical arguments regarding the market benefits of price responsive demand.
From the viewpoint of classical arguments, the demand curve as a function of price always represents an optimal solution that maximizes (aggregated) consumer utility given a budget constraint.
However, the existence of inertia of demand suggests that such inertia may result in a restriction of feasible choices in consumption space, which then prevents consumer from engaging in the optimal consumption choice predicted by classical consumer theory.
This implies that an instantly observed static demand curve may not fully reflect consumers' utility structure regardless of consumers' rationality, which may impact negatively on overall market efficiency under previously suggested pricing solutions for electricity markets.

While there may exist a \emph{nominal demand curve} well reflecting a consumer's utility, 
the inability to respond instantly creates a temporal distortion in demand curve in two respects:
(1) Since the instantaneous price elasticity of demand is near zero, an instantaneous demand curve during the time slot with a sudden price surge may be represented by a vertical line in a classical quantity-price plot, so that it may not coincide with the nominal demand curve; and
(2) The demand hedged against the risk of a repeated price surge would not respond to subsequent price reduction, so that an instantaneous demand curve after an unruffled price may not coincide with the nominal demand curve for a certain subsequent period of time.
%would give an same effect / play as if a fixed price for a temporal period of time

\subsection{The Limitations of High-frequency Real-time Pricing}

It is generally expected that DR will be beneficial both in terms of system operability as well as economical perspective, under the high penetration of VERs.
However, the distortion in the demand curve due to demand inertia can degrade the overall expected benefits from DR programs designed incautiously without consideration of demand inertia.
An important example of the latter is high-frequency RTRP.
In this subsection, we discuss the limitation of potential benefits from high-frequency RTRP under supply fluctuation caused by VERs.

\begin{figure}[!t]
\centerline{
	\includegraphics[width = 2.5in]{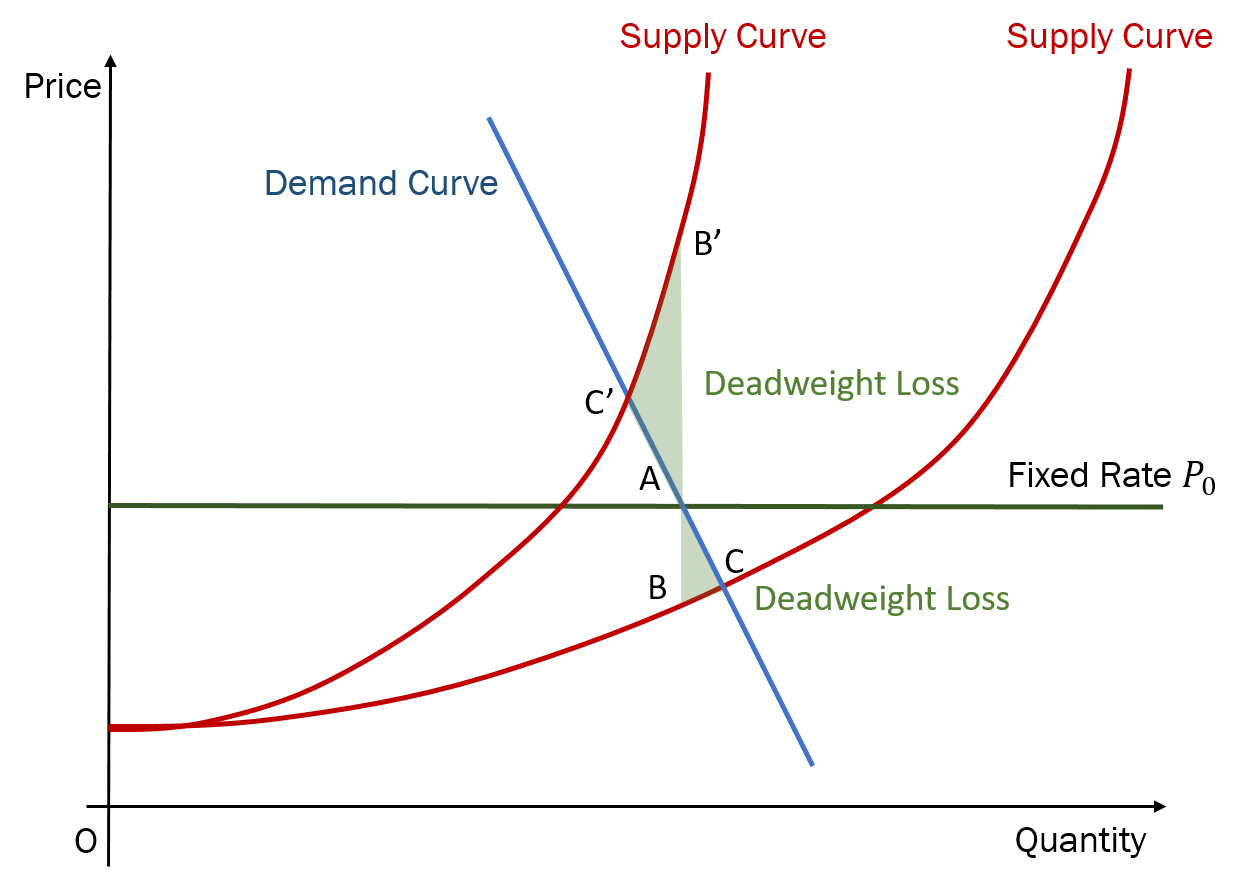}
}
\caption{The analysis of economic inefficiency resulting from a fixed retail electricity tariff under supply fluctuation by VER.}
\label{fig:EconIneffFixRateVER}
\end{figure}

While the analysis on the allocative efficiency of RTRP compared to fixed price under volatile demand analyzed in previous literature, is depicted in Figure~\ref{fig:EconIneffFixRate},  
this analysis can be extended to the allocative efficiency of RTRP compared to fixed price under supply fluctuation by VERs as depicted in Figure \ref{fig:EconIneffFixRateVER}. Analogously, the shaded triangles $\Delta ABC$ and $\Delta AB'C'$ are the deadweight loss, representing the economic inefficiency caused by the fixed rate $P_0$, or, conversely, the expected allocative efficiency benefit from RTRP compared to a fixed tariff. According to the classical arguments on RTRP, more frequent price change would be more beneficial, because they would more accurately the balance supply and demand in real time, so that it is more informative for consumers to make an optimal decision.

\begin{figure}%[!t]
\centerline{
    \subfigure[The VER drop-off case.]{
        \includegraphics[width = 1.8in]{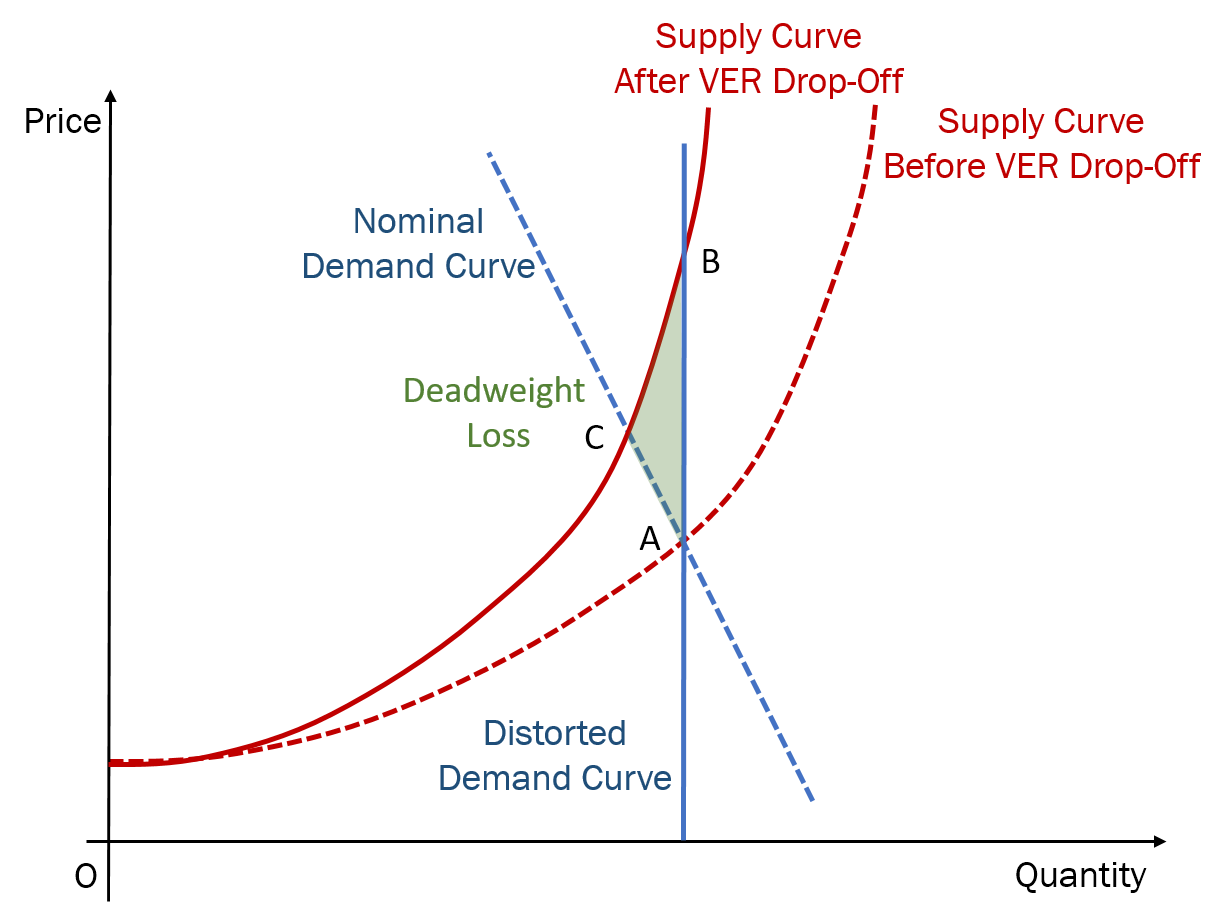}
        \label{fig:IneffVERDropOff}
    }
    \subfigure[The VER restoration case.]{
        \includegraphics[width = 1.8in]{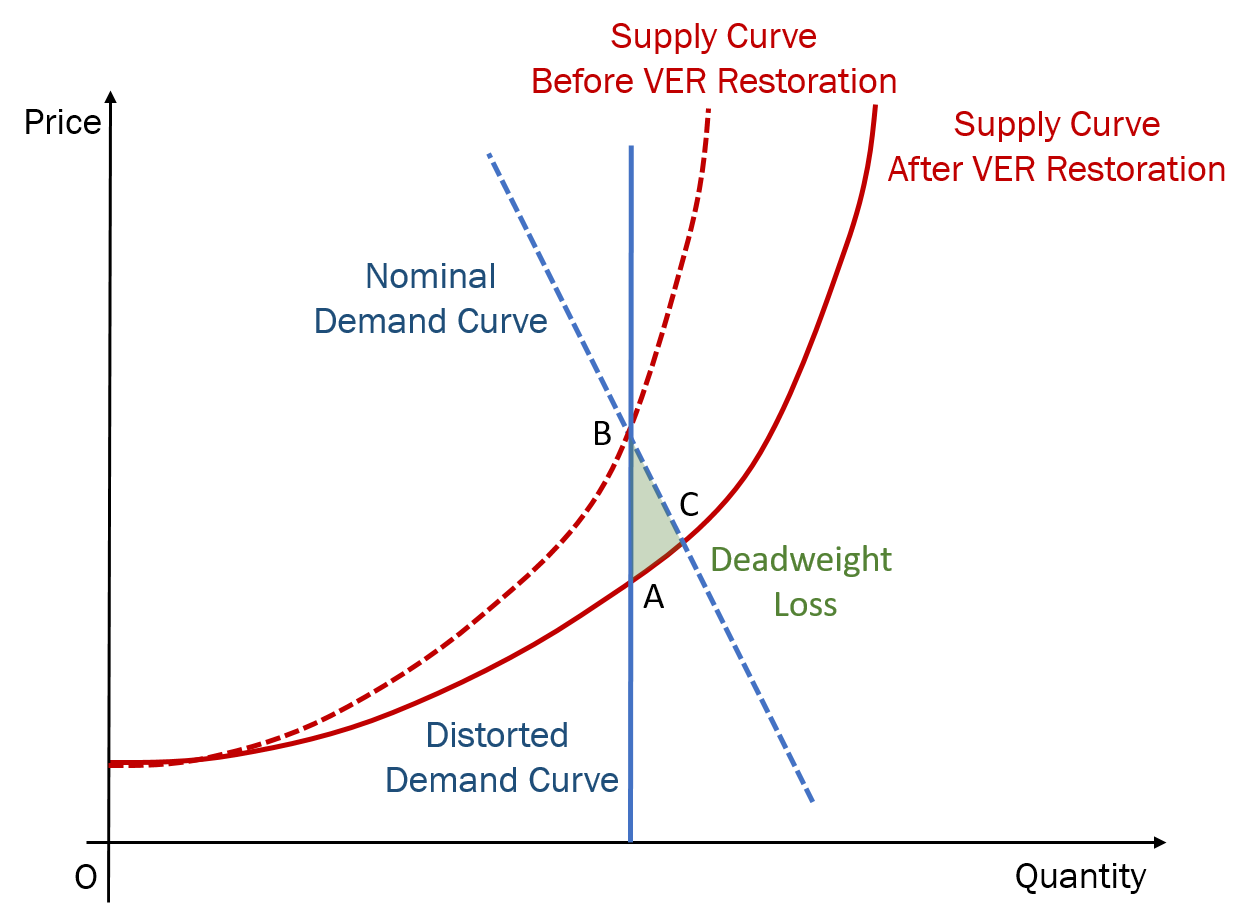}
        \label{fig:IneffVERRestoration}
    }
}
\caption{The analysis of economic inefficiency resulting from the inability of an instant demand response under supply fluctuation by VER.}
\label{fig:IneffVER}
\end{figure}

However, our observation suggests that such allocative efficiency is not likely to be achievable because of the distortion in demand curve caused by demand inertia.
The inability of customers to respond instantly may distort the demand curve to a vertical line in quantity-price plot. 
Noting this distortion, the demand behavior in practice may be realized as if it is exposed to fixed price.
The impact on allocative efficiency from the demand curve distortion is presented in Figure \ref{fig:IneffVER}.

Figure \ref{fig:IneffVERDropOff} depicts the situation where VER drops due to events such as sudden diminution of wind or cloud cover blocking the sun.
Suppose that the market equilibrium point before the VER drop is $A$.
Then, the VER drop shifts the supply curve to the left.
After the supply curve shift, the optimal market clearing point maximizing social welfare is $C$ on the nominal demand curve.
However, the actual market clearing point is realized at $B$ due to demand curve distortion. 
Hence, the shaded triangle $\Delta ABC$ is the deadweight loss, exhibiting the economic inefficiency following from demand inertia.

In Figure \ref{fig:IneffVERRestoration}, the analysis of the situation where VER is restored by incidents e.g. the increase in wind generation caused by a gust of wind, or that of solar generation following a cloud gap is provided.
Suppose the market equilibrium point before VER restoration is $A$.
The increase in generation followed by VER restoration event results in the newly formed supply curve to lie on the right hand side of the previous one.
While $C$ on the nominal demand curve is the optimal market clearing point after the supply curve shift,
the demand curve distortion caused by the hedging of demand against the risk of a repeated VER drop-off may result in the actual market clearing point being realized at $B$. 
Again, the shaded triangles $\Delta ABC$ indicates the deadweight loss, the economic inefficiency from demand inertia.

The allocative (in)efficiency analysis of RTRP under demand inertia suggests that there is a fundamental limitation to achieving market efficiency that can be expected from traditional market efficiency analysis without consideration of demand inertia. This necessitates a redefinition of market efficiency from an optimal control theoretical perspective.
In addition, the demand behavior under RTRP is as if it is exposed to a fixed price, leading to another crucial implication that RTRP may not significantly resolve the vulnerability of markets from the exercise of market power.
The differences in the ability of various market participants to control their behavior endows different market powers to each market participant; the more instantaneously responsive market participant has an advantage over market participants with larger delay. Such a combination of differentially endowed market powers makes market more vulnerable to the exercise of market power or market manipulation. A similar argument is found in financial markets with high-frequency trading (HFT) practices, in terms of the robustness with respect to market manipulation and market fairness \cite{Biais}. 
Moreover, the inability of demand response to instantaneously respond also suggests that RTRP may contribute negatively to demand volatility mitigation, so that the savings in the cost of maintaining reserve capacity may be less than expected under previous literature.

\section{Conclusion}  \label{sec:Conc}

A market is a dynamical system that is
designed to proceed toward an optimal state as its equilibrium. 
However, such a process necessarily requires a certain amount of time to reach its equilibrium. 
While dynamic modeling and control on the generation side in power systems has been well understood, the understanding of dynamic behavior on the demand side in response to price has been unclear.
In this paper, we consider a consumer's dynamic behavior in response to real time price change, 
by studying empirical data on a price-responsive load in the ERCOT area. 

Out empirical study suggests the following: 
(1) \emph{the price responsiveness of demand may exhibit qualitatively different behavior in response to ``normal price'' and ``high price'';} and 
(2) \emph{there exists a demand response delay consequent on a high price surge at peak hours.}

Such behavioral features imply that frequent price changes do not necessarily bring economic efficiency in the sense of social welfare maximization.

This idea provides important guidance for designing two fundamental factors in time-varying
retail electricity prices, \textit{frequency} and \textit{timeliness}. 
Here 
``frequency of price" is the frequency at which retail prices change, and 
``timeliness of price" is the time lag between when a price is set and when it is effective \cite{Borenstein05b}. 
It is generally assumed among economists that RTRP with high frequency and just-in-time timeliness would be ideal in terms of economic efficiency in the electricity market, 
as RTRP is an attempt to provide more accurate signals closely reflecting the
actual supply/demand status in the market. 
However, the inference based on our work is that neither argument is necessarily right. 
The inherent delay in the responsiveness of loads to high price volatility exacerbates the predictability of price, 
thereby making demand less responsive to RTRP, 
which in fact worsens economic efficiency. 
Consumers which are more exposed to market volatility stiffen their demand to be more inelastic and tend to be more conservative due to the inertial nature of demand. 
This suggests that there exists a trade-off between controllability of demand and observability of markets, 
so that there may exist an optimal frequency and timeliness which should be carefully considered for optimal pricing design. 
This also supports the importance of relatively long-term contract markets such as day-ahead electricity markets. 
Market efficiency should be re-analyzed taking into consideration the trade-off between the controllability of demand and the observability of the market.
In subsequent work, we aim to provide a rigorous analysis of consumer rationality and develop a quantitative prediction model for demand response.

\end{document}